\documentclass[12pt]{article}

\usepackage{amsfonts, amsmath}

\newcommand{\be}{\begin{equation}} \newcommand{\ee}{\end{equation}}

\begin{document}
\title{Generalized uncertainty relations in a quantum theory and thermodynamics from the uniform point of view.} \thispagestyle{empty}

\author{A.E.Shalyt-Margolin\thanks
{Phone (+375) 172 883438; e-mail
alexm@hep.by}\hspace{10pt} and \hspace{5pt} A.Ya.Tregubovich
\thanks{Phone (+375) 172 840441; e-mail a.tregub@open.by}
}
\date{}
\maketitle
 \vspace{-15pt}
{\footnotesize\noindent {\large $^*$} National Centre of High Energy and
Particle Physics Bogdanovich Str.153, Minsk \hspace*{8pt} 220040, Belarus\\
{\large $^\dagger$} Institute of Physics National Academy of Sciences
                   Skoryna av.68, Minsk\\\hspace*{8pt} 220072, Belarus}\\

%\begin{abstract}
{\bf\small\noindent Abstract}\\ {\footnotesize
 A generalization of the thermodynamic uncertainty relations is proposed.
 It is done by introducing of an additional term proportional to the interior energy
  into the standard thermodynamic uncertainty relation that leads to existence
  of the lower limit of inverse temperature.}
 \vspace{0.5cm}
{\ttfamily{\footnotesize
\\ PACS: 03.65; 05.70\\ \noindent Keywords:
                   generalized   uncertainty relations; generalized
                   uncertainty\\ relations in thermodynamics}}

\rm\normalsize \vspace{0.5cm}
%\end{abstract}

 It is well known that in thermodynamics an inequality for the pair interior energy -
 inverse temperature, which is completely analogous to the standard uncertainty
 relation in quantum mechanics can be written down \cite{r6} -- \cite{r8}. The
 only (but essential) difference of this inequality from the quantum mechanical
 one is that the main quadratic fluctuation is defined by means of
 classical partition function rather than by quantum mechanical expectation values.
 In the last 14 - 15 years a lot of papers appeared in which the  usual
 momentum-coordinate uncertainty relation has been modified at very high
 energies of order Planck energy $E_p$ \cite{r2}--\cite{r5}. In this note we
 propose simple reasons for modifying the thermodynamic uncertainty relation at
 Planck energies. This modification results in existence of the minimal
 possible main quadratic fluctuation of the inverse temperature. Of course we
 assume that all the thermodynamic quantities used are properly defined so that
 they have physical sense at such high energies.

We start with usual Heisenberg uncertainty relations \cite{r1} for momentum -
coordinate:
\begin{equation}\label{U1}
 \Delta x\geq\frac{\hbar}{\Delta p}.
\end{equation}
 It was shown that at the Planck
 scale a high-energy term must appear:
\begin{equation}\label{U2}
\Delta
x\geq\frac{\hbar}{\Delta p} + \, const\, L_{p}^2\,\,\frac{\Delta p}{\hbar}.
\end{equation}
where $L_{p}$ is the Planck length
$L_{p}^2 = G\hbar /c^3 \simeq 1,6\;10^{-35}m$.
In \cite{r2} this term is derived from the string theory, in \cite{r3}
 it follows from the simple estimates of Newtonian gravity and quantum mechanics,
 in \cite{r4} it comes from the black hole physics, other methods can also be
 used \cite{r5}. Particularly the coefficient $const$ is shown to be unity in
 paper \cite{r3}.
Relation (\ref{U2}) is quadratic in $\Delta p$
\begin{equation}\label{U4}
 L_{p}^2\, ({\Delta p})^2 - \hbar\,\Delta x\Delta p + \hbar^2 \leq0
\end{equation}
 and therefore leads to the fundamental length
\begin{equation}\label{U5}
 \Delta x_{min}=2L_{p}
\end{equation}
  Using relations (\ref{U2}) it is easy to obtain a similar relation for the
 energy - time pair. Indeed (\ref{U2}) gives
\begin{equation}\label{U6}
\frac{\Delta x}{c}\geq\frac{\hbar}{\Delta p c }+L_{p}^2\,\frac{\Delta
p}{c \hbar},
\end{equation}
then
\begin{equation}\label{U7}
\Delta t\geq\frac{\hbar}{\Delta
E}+\frac{L_{p}^2}{c^2}\,\frac{\Delta p c}{\hbar}=\frac{\hbar}{\Delta
E}+t_{p}^2\,\frac{\Delta E}{ \hbar}.
\end{equation}
where the smallness of $L_p$ is taken into account so that the difference
between $\Delta E$ and $\Delta (pc)$ can be neglected and $t_{p}$  is the
Planck time $t_{p}=L_p/c=\sqrt{G\hbar/c^5}\simeq 0,54\;10^{-43}sec$.
Inequality (\ref{U7}) gives analogously to (\ref{U2}) the lower boundary
for time $\Delta t\geq2t_{p}$ determining the fundamental time
\begin{equation}\label{U10}
 \Delta t_{min}=2t_{p}.
 \end{equation}
 Thus, the inequalities discussed can be rewritten in a standard form
\begin{equation}\label{U11}
\left\{ \begin{array}{ll}
\Delta x & \geq\frac{\displaystyle\hbar}{\displaystyle\Delta p}+
\left(\frac{\displaystyle\Delta p}{\displaystyle P_{pl}}\right)\,
\frac{\displaystyle\hbar}{\displaystyle P_{pl}} \\
 & \\
 \Delta t & \geq\frac{\displaystyle\hbar}{\displaystyle\Delta E}+
 \left(\frac{\displaystyle\Delta E}{\displaystyle E_{p}}\right)\,
 \frac{\displaystyle\hbar}{\displaystyle E_{p}}
\end{array} \right.
\end{equation}
where $P_{pl}=E_p/c=\sqrt{\hbar c^3/G}$.
 Now we
consider the thermodynamics uncertainty relations between the inverse temperature
and interior energy of a macroscopic ensemble
\begin{equation}\label{U12}
\Delta \frac{1}{T}\geq\frac{k}{\Delta U}.
\end{equation}
where $k$ is the Boltzmann constant. \\ N.Bohr \cite{r6} and
W.Heisenberg \cite{r7} first pointed out that such kind of uncertainty principle
should take place in thermodynamics. The thermodynamic
uncertainty  relations (\ref{U12})  were proved by many authors and in various
ways \cite{r8}. Therefore their validity does not raise any doubts.
Nevertheless, relation (\ref{U12}) was proved in view of the standard model of
the infinite-capacity heat bath encompassing the ensemble. But it is obvious
from the above inequalities that at very high energies the capacity of the heat
bath can no longer to be assumed infinite at the Planck scale. Indeed, the total energy of the pair heat bath -
ensemble may be arbitrary large but finite merely as the universe is born at a
finite energy. Hence the quantity that can be interpreted as the temperature
of the ensemble must have the upper limit and so does its main quadratic deviation.
In other words the quantity $\Delta (1/T)$ must be bounded from below. But in this
case an additional term should be introduced into (\ref{U12})
\begin{equation}\label{U12a}
\Delta \frac{1}{T}\geq\frac{k}{\Delta U} + \eta\,\Delta U
\end{equation}
where $\eta$ is a coefficient. Dimension and symmetry reasons give
$$
\eta = \frac{k}{E_p^2}.
$$
As in the previous cases inequality (\ref{U12a}) leads to the fundamental
(inverse) temperature.
\begin{equation}\label{U15}
T_{max}=\frac{\hbar}{2t_{p}
k}=\frac{\hbar}{\Delta t_{min} k}, \quad
\beta_{min} = {1\over kT_{max}} =  \frac{\Delta t_{min}}{\hbar}
\end{equation}
Thus, we obtain the system of generalized uncertainty relations in a symmetric
form
\begin{equation}\label{U17}
\left\{
\begin{array}{lll}
\Delta x & \geq & \frac{\displaystyle\hbar}{\displaystyle\Delta p}+
\left(\frac{\displaystyle\Delta p}{\displaystyle P_{pl}}\right)\,
\frac{\displaystyle\hbar}{\displaystyle P_{pl}} \\
&  &  \\
\Delta t & \geq & \frac{\displaystyle\hbar}{\displaystyle\Delta E}+
\left(\frac{\displaystyle\Delta E}{\displaystyle E_{p}}\right)\,
\frac{\displaystyle\hbar}{\displaystyle E_{p}}\\
  &  &  \\
  \Delta \frac{\displaystyle 1}{\displaystyle T}& \geq &
  \frac{\displaystyle k}{\displaystyle\Delta U}+
  \left(\frac{\displaystyle\Delta U}{\displaystyle E_{p}}\right)\,
  \frac{\displaystyle k}{\displaystyle E_{p}}
\end{array} \right.
\end{equation}
or in the equivalent form
\begin{equation}\label{U18}
\left\{
\begin{array}{lll}
\Delta x & \geq & \frac{\displaystyle\hbar}{\displaystyle\Delta p}+
L_{p}^2\,\frac{\displaystyle\Delta p}{\displaystyle\hbar} \\
  &  &  \\
  \Delta t & \geq &  \frac{\displaystyle\hbar}{\displaystyle\Delta E}+
  t_{p}^2\,\frac{\displaystyle\Delta E}{\displaystyle\hbar} \\
  &  &  \\
  \Delta \frac{\displaystyle 1}{\displaystyle T} & \geq &
  \frac{\displaystyle k}{\displaystyle\Delta U}+
  \frac{\displaystyle 1}{\displaystyle T_{p}^2}\,
  \frac{\displaystyle\Delta U}{\displaystyle k}
\end{array} \right.
\end{equation}
Here $T_{p}$ is the Planck temperature: $T_{p}=\frac{E_{p}}{k}$. \\
In the conclusion we would like to note that the restriction on the heat bath
made above turns the equilibric  partition function to be non-Gibbsian
\cite{r9}.


\begin{thebibliography}{99}
%
%
\bibitem{r1}
W.Heisenberg,Zeitsch.fur Phys.v.43(1927)P.172
%
%
\bibitem{r6}
N.Bohr Faraday Lectures pp. 349-384, 376-377 Chemical
Society, London (1932)
%
%
\bibitem{r7}
W.Heisenberg Der Teil und Das Ganze ch 9 R.Piper, Munchen (1969)
%
%
\bibitem{r8}
J.Lindhard
Complementarity between energy and temperature In:
The Lesson of Quantum Theory Ed. by J. de Boer, E.Dal and O.Ulfbeck
North-Holland, Amsterdam (1986);\\
 Lavenda Statistical Physics: a
Probabilistic Approach J.Wiley and Sons, N.Y. (1991);\\
B.Mandelbrot, IRE Trans. Inform. Theory IT-2 (1956) 190;\\
L.Rosenfeld In: Ergodic theories Ed. by P.Caldrirola Academic
Press, N.Y. (1961);\\
F.Schlogl J. Phys. Chem. Solids 49 (1988) 679;\\
J.Uffink and J. van Lith-van Dis Found. of Phys. 29 (1999) 655
%
%
\bibitem{r2}
G.Veneziano Europhys.Lett 2 (1986) 199;\\
D.Amati, M.Ciafaloni and G.Veneziano
Phys.Lett. B197 (1987) 81; Ibid. B216 (1989) 41
%
%
\bibitem{r3}
R.Adler Mod.Phys.Lett. A14 (1999) 1371, gr-qc/9904026;\\
M.Maggiore Phys.Lett. B304 (1993) 65, hep-th/9301067
%
%
\bibitem{r4}
M.Maggiore Phys.Lett. B304 (1993) 65, hep-th/9301067;\\
M.Maggiore hep-th/9310157
%
%
\bibitem{r5}
M.Maggiore Phys.Rev. D49 (1994) 5182, hep-th/9305163;\\
M.Maggiore Phys.Lett. B319 (1993) 83, hep-th/9309034;\\
S.Capozziello, G.Lambiase, and G.Scarpetta gr-qc/9910017
%
%
\bibitem{r9}
F.Pennini,A.Plastino, and A.R.Plastino cond-mat/0110135
%
%
\end{thebibliography}
\end{document}